\documentclass[aps,prd,a4paper,nofootinbib,
preprintnumbers,superscriptaddress,showpacs,showkeys,twocolumn]{revtex4-2}

\usepackage{amsmath}
\usepackage{amssymb}
\usepackage{bm}
\usepackage{graphicx}
\usepackage{color}

\newcommand{\be}{\begin{equation}}
\newcommand{\ee}{\end{equation}}
\newcommand{\ba}{\begin{eqnarray}}
\newcommand{\ea}{\end{eqnarray}}

\begin{document}

\title{Anisotropic fluctuations of 
	angular momentum of heavy quarks in the Glasma}

\author{Pooja}
\email{pooja19221102@iitgoa.ac.in}
\affiliation{School of Physical Sciences, Indian Institute of Technology Goa, Ponda-403401, Goa, India}

\author{Santosh K. Das}
\email{santosh@iitgoa.ac.in}
\affiliation{School of Physical Sciences, Indian Institute of Technology Goa, Ponda-403401, Goa, India}

\author{Vincenzo
	 Greco}\email{greco@lns.infn.it}
\affiliation{Department of Physics and Astronomy "Ettore Majorana", University of Catania, Via S. Sofia 64, I-95123 Catania, Italy}
\affiliation{INFN-Laboratori Nazionali del Sud, Via S. Sofia 62, I-95123 Catania, Italy}

\author{Marco Ruggieri}\email{marco.ruggieri@dfa.unict.it}
\affiliation{Department of Physics and Astronomy "Ettore Majorana", University of Catania, Via S. Sofia 64, I-95123 Catania, Italy}

\begin{abstract}
We study the evolution of the angular 
momentum of the heavy quarks in 
the very early stage of high energy nuclear collisions,
in which the background is made of evolving
Glasma fields. Given the novelty of the
problem,
we limit ourselves to the use of
toy heavy quarks with a
large, unphysical mass, in order to implement
the kinetic equations for the angular momentum
in the non-relativistic limit. We find that 
as a consequence of the anisotropy of the 
background fields, angular momentum fluctuations are
also anisotropic: we understand this in simple terms
relating the fluctuations of the angular momentum, $\bm L$,
to those of linear momentum. While orbital angular
momentum diffuses and develops substantial fluctuations
and anisotropies, the spin does not. Hence,
we can identify the fluctuations of $\bm L$ with those
of the total angular momentum $\bm J=\bm L + \bm S$.
Therefore, our study suggests that the total angular
momentum of the heavy quarks in the early stage
of high energy nuclear collisions will present
anisotropic fluctuations. 
\end{abstract}



\maketitle

\section{Introduction}
To study the fundamentals of nature, to be precise in the context of our work, Quantum 
Chromodynamic (QCD) matter, high-energy nuclear collisions have worked as great tools. The ultrarelativistic 
heavy-ion collisions performed at Relativistic Heavy Ion Collider (RHIC) and Large Hadron Collider (LHC) made it possible to recreate 
conditions similar to those of the early universe. The term little-bang 
that is often used to describe the shattering of two relativistic nuclei confirms the formation of a deconfined state of quarks and gluons known as Quark-Gluon Plasma (QGP) ~\cite{Shuryak:2004cy,Jacak:2012dx}.
The creation of locally equilibrated QGP in these collision experiments is a consequence of very complicated dynamics that happens to occur within a time scale of 1 fm/c. 
The very earliest phase is the span of highly non-equilibrium gluon fields and we call it the initial condition for the thermalized QGP. 
This initial condition, famously known in the literature as Glasma, is recognized as the pre-equilibrium stage just after the collision of nuclei where the density of chromodynamic fields is extremely high. 
The research of this initial condition and its decay to a nearly perfect fluid QGP is of immense interest to physicists worldwide to explore the QCD phase diagram in more detail.

The scenario before the collision is described using effective theory of Color-Glass Condensate (CGC) ~\cite{McLerran:1993ni,McLerran:1993ka,McLerran:1994vd,Gelis:2010nm,Iancu:2003xm,McLerran:2008es,Gelis:2012ri} which leads to the formation of Glasma ~\cite{Kovner:1995ja,Kovner:1995ts,Gyulassy:1997vt,Lappi:2006fp, Krasnitz:2003jw,Fukushima:2006ax,Fujii:2008km,Fukushima:2013dma,Romatschke:2005pm, Romatschke:2006nk,Fukushima:2011nq}. CGC is the description of high-energy partons in the saturation regime. The two colliding nuclei are modeled as two sheets of colored glass where the fast parton dynamics seem to stop and they act as static sources for low momentum gluons. As a result of collision of two CGC sheets, we get a configuration of strong classical gluon fields, namely Glasma. The Glasma consists of longitudinal color-electric and color-magnetic fields in the weak coupling regime and is characterized by a large gluon occupation number. The evolution of these fields is studied using classical Yang-Mills (CYM) equations. We are interested in the physics of the non-equilibrium phase when the dynamics is of dense chromodynamic fields rather than partons.   
In this article, we use the notation EvGlasma for the evolving Glasma fields and reserve the name Glasma for the initial condition.

Heavy quarks ~\cite{Prino:2016cni, Andronic:2015wma, Rapp:2018qla, Cao:2018ews, Aarts:2016hap, Greco:2017rro, Dong:2019unq, Xu:2018gux, Moore:2004tg, vanHees:2005wb, vanHees:2007me, Gossiaux:2008jv, He:2011qa, Prakash:2021lwt, Song:2015sfa, Alberico:2011zy, Lang:2012cx,Das:2015ana, Xu:2017obm, Cao:2016gvr, Das:2016cwd, Das:2017dsh, Das:2015aga,Song:2019cqz, Beraudo:2015wsd,Das:2013kea,Berrehrah:2013mua, Scardina:2017ipo} produced in the very early phase of the ultra-relativistic heavy-ion collisions are efficient probes each for the pre-equilibrium Glasma phase and the equilibrated quark-gluon plasma (QGP). The charm and the beauty quarks are the most interesting quarks due to their very small formation time computed by $\tau_{form} \approx 1/(2m)$, where $m$ is the mass of the heavy quark. 
It gives $\tau_{form} \leq 0.08$ fm/c for charm 
quarks and 
$\tau_{form} \leq 0.03$ fm/c
for beauty quarks.
Hence, due to their large masses, they are formed immediately just after the collision, so, they can propagate in the evolving gluonic medium to probe their evolution.
Along with that, heavy quarks carry negligible color current and rarely interact among themselves due to their small number and large mass. So, they provide no disturbance to the evolving gluon fields
and behave as ideal probes of these fields
in the early stages of high energy collisions.

The prime aim of this article is to study
the behaviour of the angular momentum of 
the heavy quarks
(HQs) 
in presence of the coherent gluon fields
that form in the early stage  of the
high energy nuclear collisions, and that 
approximately live up to $0.6$ fm/c
in the case of collisions at the RHIC energy
and up to $\approx 0.4$ fm/c for collisions
at the LHC energy.
For simplicity, we limit ourselves to the case
of non-relativistic heavy quarks,
leaving the full relativistic 
case to a future study. 
It is already known that the coherent 
gluon fields of the evolving Glasma affect the
diffusion of momentum of HQs~\cite{Mrowczynski:2017kso,Ruggieri:2018rzi,Sun:2019fud,Liu:2019lac,Boguslavski:2020tqz,Liu:2020cpj,
	Khowal:2021zoo,Ipp:2020nfu}; 
it is therefore of a certain interest to 
extend the subject of the previous studies and analyze
the behavior of other quantities: among them,
we believe that angular momentum 
spreading of HQs is potentially
interesting. 
Since ours is a first study about the subject,
we limit ourselves to simulate the 
system in a static box; moreover,
the system we simulate has $\langle \bm J\rangle=0$
both in the gluon and in the HQs sectors:
the inclusion of a finite $\langle \bm J\rangle$
is feasible but far from being trivial and will be
the subject of future studies.
Our main purpose here is to show how
the anisotropic momentum diffusion of HQs in the
evolving Glasma fields leads to anisotropic
fluctuations of the angular momentum of HQs.

The plan of the article is as follows: 
in section \ref{sec:formalism}, we
briefly present the 
initial condition, namely the Glasma fields, 
as well as their evolution, and write the kinetic
equations for the HQs. 
In section \ref{sec:results}, 
we present our results for the 
diffusion of the total angular
momentum of HQs in the evolving Glasma. 
Finally, we summarize our results
presenting the anisotropy of  
the fluctuations of the
components of $\bm J$. In 
section \ref{sec:conclusions},  we conclude our work 
and discuss the possible future improvements.

\section{Formalism\label{sec:formalism}}
\subsection{Notations and conventions \label{subsec:notations}}
We work in natural units, explicitly, $c=\hbar=k_B=1$.
The Greek indices such as $\mu$, $\nu$ can take the values 0, 1, 2, 3; where 0 represents the temporal coordinate and 1, 2, 3 represent the spatial coordinates. On the other hand, the Latin alphabets, namely $i$, $j$, $k$ numerate the Cartesian spatial coordinates and can take values 1, 2, 3 only. Keeping Einstein's summation convention in mind, we assume repeated indices are summed up. $\varepsilon_{ijk}$ denotes the three dimensional Levi-Civita tensor with $\varepsilon_{123} = \varepsilon^{123} =+1$.

The indices $a$, $b$, $c$ having values $1,2,3,...N_c^2-1$ are reserved for the color components of the $SU(N_c)$ group in the adjoint representation.
$T^a$'s are the generators of the group following the normalisation condition $Tr(T^aT^b)=\delta^{ab}/2$ and the commutation relation $[T^a, T^b] = if^{abc} T^c$ where $f^{abc}$ are the completely anti-symmetric structure constants.

The QCD covariant derivative is
\begin{equation}
	\mathcal{D}^\mu = \partial^\mu - i g A^\mu,
\end{equation}
This allows us to define the kinematic 4-momentum
\begin{equation}
	P^\mu = i \mathcal{D}^\mu = i\partial^\mu + g A^\mu = p^\mu+
	g A^\mu.
\end{equation}
where $p^\mu$ denotes the canonical momentum.
This gives in particular
\begin{equation}
	P^i = i\partial^i + g A^i=
	-i\frac{\partial}{\partial x^i}+ g A^i=
	-i\nabla_i + g A^i,\label{eq:three_1}
\end{equation}
where \begin{equation}
	\nabla_i = \frac{\partial}{\partial x^i}.
\end{equation}
Using 3-dimensional notation, we can rewrite
Eq.~\eqref{eq:three_1} as
\begin{equation}
	\bm P =\bm p + g\bm A =
	-i\bm\nabla + g \bm A.\label{eq:three_1_2}
\end{equation}


\subsection{Glasma and its evolution}
The 
initialization of the gluon fields we adopt in 
our study is based on
the McLerran-Venugopalan (MV) model~\cite{McLerran:1993ni,McLerran:1993ka,McLerran:1994vd}
in which  
the collision of high energy nuclei 
is described within the framework of 
the color-glass condensate (CGC). 
CGC is an effective field theory based on the separation of scales of nucleon momentum fraction. The two colliding objects are viewed as two thin Lorentz-contracted sheets of a colored glass. CGC enables us to treat the two degrees of freedom, i.e., very fast color sources and slow color fields on different footing. The fast partons, as a result of time dilation appear to be frozen and act as the sources of the slow dynamical gluon fields in the saturation regime. These dominating gluon fields behave classically due to their high occupation numbers.  

As per the MV model, the color charge densities of fast partons vary randomly on nuclei. Hence, it is assumed that the static color charge densities $\rho^a_{A}$ on colliding entity $A$ are normally distributed random variables. The first and second moments of the color charge density are given by
\begin{eqnarray}
	\langle \rho^a_{A}(\textbf{\textit{x}}_T)\rangle &=& 0,
	\label{eq:colorchargedensity_1}\\
	\langle \rho^a_{A}(\textbf{\textit{x}}_T)\rho^b_{A}(\textbf{\textit{y}}_T)\rangle &=& (g\mu_A)^2   \delta^{ab} \delta^{(2)}(\textbf{\textit{x}}_T - \textbf{\textit{y}}_T), \label{eq:colorchargedensity_2}
\end{eqnarray}
where $g$ is the Yang-Mills coupling constant and $g\mu_A$ denotes the color charge density of object $A$ in the transverse plane, which is of the order of the saturation momentum $Q_s$; 
$a$ and $b$ correspond to the adjoint color index. For the sake of simplicity, we
limit ourselves to the $SU(2)$ color group, 
hence, we get $a, b=1,2,3$.

Within the MV model, the non-abelian interaction
of the color fields present in the nuclei
before the collision at $t=0^-$ 
produces a new set of strong,
longitudinal fields after the collision at $t=0^+$, 
that serves
as the initial condition for the problem at hand:
this new set of fields is called the Glasma.
In order to
determine the Glasma, 
we start by solving the two-dimensional Poisson's equation for the gauge potential generated by the static color charge distribution of the $m^{th}$ nuclei,
	\begin{equation}
		-\partial^2_{\perp}\Lambda^{(m)}(\textbf{\textit{x}}_T)
		=\rho^{(m)}(\textbf{\textit{x}}_T),
	\end{equation}
where $m=A, B$ are the two
colliding objects. 
The corresponding Wilson lines are computed by
\begin{eqnarray}
V^\dagger(\textbf{\textit{x}}_T) &=& e^{-ig\Lambda^{(A)}(\textbf{\textit{x}}_T)},\\
W^\dagger(\textbf{\textit{x}}_T) &=& e^{-ig\Lambda^{(B)}(\textbf{\textit{x}}_T)}.
\end{eqnarray}
The Wilson lines are used
 to calculate the transverse components of the gauge fields present on the colliding object. The pre-collision gauge fields are given by 
\begin{eqnarray}
	\alpha_i^{(A)} &=& \frac{-i}{g}V\partial_iV^\dagger,
	\label{eq:alpha1pp}\\
	\alpha_i^{(B)} &=& \frac{-i}{g}
	W\partial_iW^\dagger.\label{eq:alpha2pp}
\end{eqnarray}
 where $i = x, y$. We assume the z-direction to be the direction of receding of two nuclei.
In terms of these gauge fields, the solution of the classical Yang-Mills (CYM) equations in the forward light cone at initial time, namely the Glasma gauge potentials, are given by 
\begin{eqnarray}
	A_i &=& \alpha_i^{(A)} +\alpha_i^{(B)},\\
	A_z &=& 0.
\end{eqnarray}	
Hence, the initial longitudinal Glasma fields are calculated by	
\begin{eqnarray}
	E^z &=& -ig \sum_{i=x,y}\bigg[\alpha^{(B)}_i, \alpha^{(A)}_i\bigg],\\
	B^z &=& -ig \bigg(\bigg[\alpha^{(B)}_x, \alpha^{(A)}_y\bigg]+\bigg[\alpha^{(A)}_x, \alpha^{(B)}_y\bigg]\bigg);
\end{eqnarray}
the transverse fields are absent in the initial condition.

After settling down the initial conditions for the Glasma, we study its evolution 
by virtue of the 
classical Yang-Mills (CYM) equations. 
We work in the temporal gauge $A_0=0$, which leaves us with the freedom to perform time-independent gauge rotations.	
In this gauge, the Hamiltonian density is given by \cite{Kunihiro:2010tg}
\begin{equation}
H = \frac{1}{2}E^a_i(x)^2 + \frac{1}{4}F^a_{ij}(x)^2.	              
\end{equation}
The field strength tensor is given by
	\begin{equation}
		F^a_{\mu\nu}(x) = \partial_\mu A^a_\nu(x)- \partial_\nu A^a_\mu(x)+ g f^{abc} A^b_\mu(x)A^c_\nu(x),
	\end{equation}
where $f^{abc} = \varepsilon^{abc}$ for $SU(2)$ gauge
theory. 
Hence, the equations of motion for the dynamical evolution of Glasma, namely the CYM equations become
\begin{eqnarray}
	&&	\frac{dA^a_i(x)}{dt}= E^a_i(x),\\
	&&	\frac{dE^a_i(x)}{dt}= \partial_jF^a_{ji}(x)+g f^{abc} A^b_j(x)F^c_{ji}(x).
\end{eqnarray}

\subsection{Evolution of heavy quarks in the
classical, non-relativistic limit}
We now quickly discuss the derivation of the
non-relativistic equations of motion of the
HQs in the evolving Glasma fields.
This derivation is pretty standard 
and was presented years ago
in the relativistic case~\cite{Heinz:1984yq},
therefore here we limit ourselves to show only
a few key steps of the derivation.

The non-relativistic Dirac equation for a free
HQ reads
\begin{equation}
	H_\mathrm{NR}\xi=E_\mathrm{NR}\xi,\label{eq:acca_1}
\end{equation}
where $\xi$ denotes the upper component of the Dirac spinor and 
\begin{equation}
	H_\mathrm{NR}=\frac{(\bm\sigma\cdot\bm p)^2}{2m},
	\label{eq:acca_2}
\end{equation}
and $m$ is the heavy quark mass and $E_\mathrm{NR}=\sqrt{\bm p^2+m^2}-m$ is the non-relativistic
energy eigenvalue.

As in the case of the interaction with an external electromagnetic
field, we can account for the interaction with a gluon field 
by the replacement $\bm p \rightarrow \bm p + g \bm A
= \bm p + g \bm A_a T_a$ in 
\eqref{eq:acca_2}, in agreement with gauge invariance,
see Eq.~\eqref{eq:three_1_2}. This gives,
in the gauge $A_0=0$,
\begin{equation}
	H_\mathrm{NR}=\frac{[\bm\sigma\cdot (\bm p+g\bm A)]^2}{2m}.
	\label{eq:acca_2a}
\end{equation}
We can extract the interaction of spin with the color-magnetic field from the operator on the right hand side of Eq.~\eqref{eq:acca_2a} as follows.
Firstly, we note that $\sigma^i\sigma^j=\delta^{ij}+
i\varepsilon_{ijk}\sigma^k$, and using $\varepsilon_{ijk}p^i p^j=0$, we can write
\begin{eqnarray}
	H_\mathrm{NR}\xi &=&
	\frac{1}{2m}
	(\bm p+g\bm A)^2\xi
	\nonumber\\
	&& + ig\varepsilon_{ijk}\sigma^k(p^i A^j
	+A^i p^j + g A^i A^j)
	 \xi.
	\label{eq:acca_3_gg}
\end{eqnarray}
The first addendum on the right hand side of the above equation
describes the diamagnetic interaction with the external field
and does not need further manipulations.  The second addendum
represents the paramagnetic interaction.

Following the same, well-known steps used to derive the non-relativistic limit of the Dirac equation in an external electromagnetic field, we get
\begin{equation}
	H_\mathrm{NR}\xi=	\left[\frac{(\bm p+g\bm A)^2}{2m}	-\frac{g}{4m}\varepsilon_{ijk} F_a^{ij}T_a\sigma^k	\right]\xi,
	\label{eq:acca_3_ggaA}
\end{equation}
where the magnetic part of field strength tensor is defined as
\begin{equation}
		F_a^{ij}=\partial^i A^{j}_a - \partial^j A^{i}_a + g f_{abc}A^{i}_b A^{j}_c.
\end{equation}
Putting
\begin{equation}
B^k_a = -\frac{1}{2}\varepsilon_{ijk}F^{ij}_a,~~~	B^k = B^k_aT_a,
\end{equation}
we get eventually
\begin{equation}
	H_\mathrm{NR}\xi=
	\left[\frac{(\bm p+g\bm A)^2}{2m}
	+\frac{g}{m} \bm S \cdot \bm B
	\right]\xi,
	\label{eq:acca_3_ggaF}
\end{equation}
with $\bm S = \bm\sigma/2$.
We notice that the operator $\bm S \cdot \bm B$ is not diagonal
in color space, due to the $T_a$ matrices in $\bm B$. 
This implies that
in general
the eigenstates of $T_3$ and $T_8$ are not eigenstates of
$H_\mathrm{NR}$, therefore time evolution will mix the different
colors.  

The classical equations of motion of momentum,
angular momentum and color charge can be obtained
easily starting with the Hamiltonian in 
Eq.~\eqref{eq:acca_3_ggaF}, by deriving the relevant
Heisenberg equations then writing them in the
classical limit. This is an easy exercise that
was done already in the relativistic case~\cite{Heinz:1984yq},
hence we can limit ourselves to quote the final
results, namely
\begin{eqnarray}
\frac{d  x^i  }{dt} &=& 	\frac{P^i}{m},	\label{eq:acca_91_3naMF_2_ar_X}\\
\frac{d  P^i  }{dt} &=&	-\frac{g}{m} Q_a \bigg(mF^{i0}_a  - {P^j}F_a^{ij} \bigg)\nonumber\\
&&	- \frac{g}{m}Q_a S^j( \mathcal{D}^{i}B^{j}_a),\label{eq:acca_91_3naMF_2_ar}\\
\frac{dQ_a}{dt}	&=&	\frac{g}{m} f_{abc} Q_c \bigg(  P^i A^{i}_b	+ S^i B^{i}_b \bigg),
	\label{eq:Ta_evolu_2_ar}\\
\frac{d S^i}{dt}&=&-\frac{g }{m}Q_aF_a^{ij}S^j,	\label{eq:spin_evol,_itisthereCL_ar}\\
\frac{d  L^i  }{dt} &=& \varepsilon_{ijk}x^j\frac{dP^k}{dt}.
\label{eq:acca_91_3naMF__ama_2}
\end{eqnarray} 
Here $Q_a=-\langle T_a\rangle$, with
$a=1,\dots,N_c^2-1$, denotes the classical
quark color charge, where $\{T_a\}$ 
stand for the
$SU(N_c)$ generators normalized as
$\mathrm{Tr}(T_a T_b)=\delta_{ab}/2$.
Equations \eqref{eq:Ta_evolu_2_ar} and \eqref{eq:spin_evol,_itisthereCL_ar} 
imply $Q_a dQ_a/dt=0$ and $S^i dS^i/dt=0$;
therefore, they describe the motion of the color charge and of the spin of the heavy quarks
on the hypersphere $\sum_a Q_a^2=\mathrm{constant}$ and on the sphere $\sum_i S_i^2=\mathrm{constant}$ respectively.
Moreover, the right hand side of
Eq.~\eqref{eq:acca_91_3naMF__ama_2} corresponds
to the torque of the force acting on the HQs.
This set of equation will be used to study the
evolution of the HQs in the background of the 
evolving Glasma fields.
In the relativistic regime the equations are 
more complicated; in particular, 
Eq.~\eqref{eq:spin_evol,_itisthereCL_ar} becomes
a BMT-like equation \cite{Bargmann:1959gz}
with terms, of order $1/m^2$,
that couple spin to $4-$momentum  \cite{Heinz:1984yq}:
this particular term does not appear in the 
non-relativistic limit as it is subleading in the
$m\rightarrow\infty$ limit. Hence, our equations
give the complete dynamics of the HQs in the
non-relativistic limit.


\subsection{Anisotropic fluctuations of $\bm L$
\label{sec:afam}}
The main scope of this study is to highlight
the anisotropic fluctuations of angular momentum
that could form in the very early stage of 
high energy nuclear collisions. 
The system we study is characterized
by a vanishing average angular momentum,
namely $\langle\bm L\rangle =0$,
both for the HQs and the background gluon fields;
however, fluctuations of $\bm L$
of the HQs are produced
as a result of the interaction with the gluon fields.
These fluctuations are anisotropic
and in fact they can be related
in a simple,
semi-quantitative way 
to the anisotropy of the momentum broadening
of HQs and to the geometry of the collision.

As a matter of fact,
$L_x=y p_z - z p_y$ which gives
\begin{equation}
	\langle L_x^2\rangle=
	\langle (y p_z - z p_y)^2\rangle
	\approx y^2 \langle p_z^2\rangle +
	z^2 \langle p_y^2\rangle,
	\label{eq:an_1}
\end{equation}
where the last equality stands in the assumption
that the components of $\bm p$ are uncorrelated
(this hypothesis is confirmed by a direct
calculation in our simulations); $y$ and $z$ 
correspond to the coordinates of the quark,
which are uncorrelated with momenta (this assumption
breaks down if $\langle \bm L\rangle\neq 0$).
Similarly we can write
\begin{equation}
	\langle L_z^2\rangle=
	\langle (x p_y - y p_x)^2\rangle
	\approx x^2 \langle p_y^2\rangle +
	y^2 \langle p_x^2\rangle.
	\label{eq:an_2}
\end{equation}
Assuming that HQs are distributed uniformly in the
box, and denoting by $\ell$ and $\ell_z$ the transverse
and longitudinal half-sides of the box respectively,
we can integrate Eqs.~\eqref{eq:an_1} and~\eqref{eq:an_2}
over the entire box and divide by the volume of the
box, $V=8\ell^2 \ell_z$; then we get
\begin{eqnarray}
	\langle L_x^2\rangle &=&  
	\frac{\ell^2}{3}
	\langle p_z^2\rangle + \frac{\ell_z^2}{3}
	\langle p_y^2\rangle ,\label{eq:an_3}\\
	\langle L_z^2\rangle &=&   
	\frac{\ell^2}{3}
	\langle p_y^2\rangle +  \frac{\ell^2}{3}
	\langle p_x^2\rangle.\label{eq:an_4}
\end{eqnarray}
Taking into account that $\langle p_y^2\rangle
= \langle p_x^2\rangle$ (the transverse momentum
distribution is isotropic, as confirmed by
the numerical simulations) and that
$\langle p_y^2\rangle + \langle p_x^2\rangle
= \langle p_T^2\rangle$,
we get
\begin{eqnarray}
	\langle L_z^2 -  L_x^2\rangle &=&   
	\frac{\ell^2}{3}
	\langle p_T^2\rangle -
	\frac{\ell^2}{3}
	\langle p_z^2\rangle - \frac{\ell_z^2}{6}
	\langle p_T^2\rangle,\label{eq:an_5}\\
	\langle L_z^2 +  L_x^2\rangle &=&  
	\frac{\ell^2}{3}
	\langle p_T^2\rangle 
	+
	\frac{\ell^2}{3}
	\langle p_z^2\rangle + \frac{\ell_z^2}{6}
	\langle p_T^2\rangle.\label{eq:an_6}
\end{eqnarray}
We thus define the anisotropy parameter
\begin{equation}
	\Delta_2 \equiv\frac{\langle L_z^2 -  L_x^2\rangle}
	{\langle L_z^2 +  L_x^2\rangle};
	\label{eq:an_77}
\end{equation}
by virtue of Eqs.~\eqref{eq:an_5}
and~\eqref{eq:an_6} we can write
\begin{equation}
	\Delta_2 =
	\frac{ 
		\langle p_T^2\rangle -
		\langle p_z^2\rangle -  
		\langle p_T^2\rangle\ell_z^2/2\ell^2}
	{\langle p_T^2\rangle +
		\langle p_z^2\rangle +  
		\langle p_T^2\rangle\ell_z^2/2\ell^2}.\label{eq:an_78}
\end{equation}

\begin{figure}[t!]
	\begin{center}
		\includegraphics[scale=0.28]{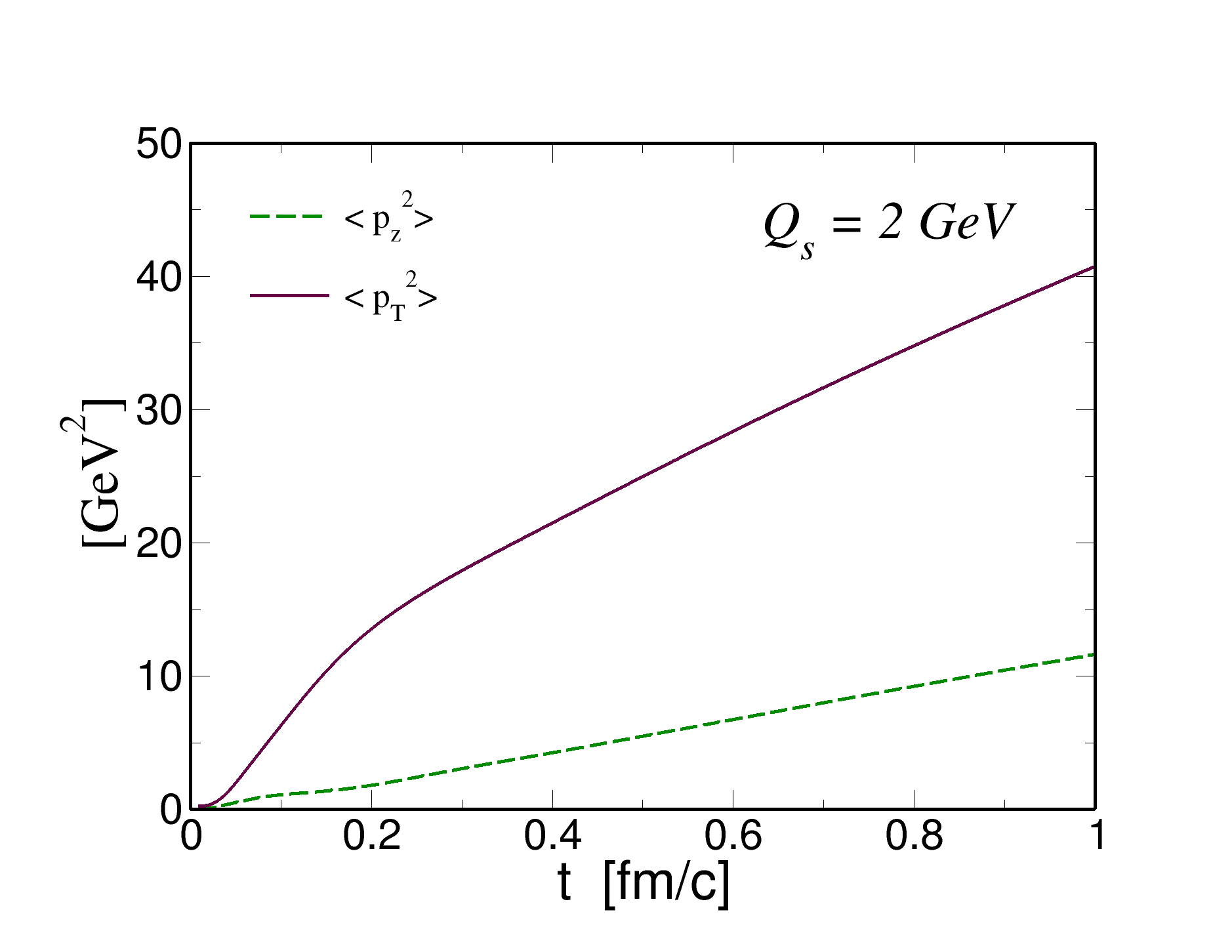}
	\end{center}
	\caption{$\langle p_z^2\rangle$ and
		$\langle p_T^2\rangle$ 
		versus time, for $Q_s=2$ GeV.\label{Fig:moman}}
\end{figure} 

The result~\eqref{eq:an_78} shows that
$\Delta_2$ can be nonzero either 
as a result of the interaction of the HQs with the
anisotropic gluon fields, which 
gives $\langle p_T^2\rangle
- \langle p_z^2\rangle>0$
as shown in Fig.~\ref{Fig:moman}, 
or because of the geometry of the fireball
created in the collision that in general,
and in particular in the pre-equilibrium stage, 
satisfies
$\ell_z \neq \ell$; in the case of the evolving Glasma
fields both conditions contribute to
$\Delta_2$.
In the limit
$\ell\gg\ell_z$, which is appropriate for the early
stage of heavy nuclei, we get
\begin{equation}
	\Delta_2 \approx
	\frac{ 
		\langle p_T^2\rangle -
		\langle p_z^2\rangle }
	{\langle p_T^2\rangle +
		\langle p_z^2\rangle }.\label{eq:an_78LN}
\end{equation} 
In this case, the dependence of $\Delta_2$ on $\ell$
and $\ell_z$ disappears.

In order to see how the geometry
affects $\Delta_2$, it is useful to study
this quantity in the case of isotropic
momentum broadening 
$\langle p_z^2\rangle = \langle p_x^2\rangle = \langle p_y^2\rangle = \langle p_T^2\rangle/2$. In this limit,
we get from Eq.~\eqref{eq:an_78}
\begin{equation}
	\Delta_2 =\Delta_2^\mathrm{geom} 
	\equiv \frac{\ell^2 - \ell_z^2}{3\ell^2 + \ell_z^2};
	\label{eq:lalonga}
\end{equation}
the superscript reminds us that Eq.~\eqref{eq:lalonga}
stands in the isotropic momentum case, 
where $\Delta_2$ takes
contribution from the geometry only.
For collisions involving large nuclei $\ell \gg\ell_z$
in the early stage (where the Glasma picture makes
sense) hence
\begin{equation}
	\Delta_2^\mathrm{geom} = \frac{1}{3},~~~
	\mathrm{large~nuclei}.
	\label{eq:unterzo}
\end{equation}
We note that if momentum broadening was isotropic,
$\Delta_2\neq 0$ because of the geometry of the
fireball, but
any deviation from Eqs.~\eqref{eq:lalonga}
can be attributed to the
anisotropic momentum broadening of the HQs.

The relevant values of $\ell$ and
$\ell_z$ in Eqs.~\eqref{eq:an_78}
and~\eqref{eq:lalonga}
can be estimated as follows:
the evolving Glasma lifetime 
is in the range of $0.3-0.6$ fm/c for 
nucleus-nucleus collisions
at the RHIC and the LHC; in this time range
the longitudinal extension 
of the fireball in the lab frame is thus 
$=O(1~\mathrm{fm})$. Hence, 
we can reasonably estimate 
the maximum $\ell_z
\approx 0.5$ fm 
in this problem.
On the other hand, $\ell$ measures the 
transverse extension of the fireball: 
for nucleus-nucleus
collisions this can be as large as $12$ fm, while
for $pA$ and $pp$ collisions this can be estimated
to be of the order of the proton radius
which is $\ell\approx 1$ fm. For the sake
of concreteness we will consider simulations for
$\ell=0.5,1$ and $2$ fm, hence limiting
ourselves to values of $\ell$
that might be appropriate 
either for the $pA$ collisions or 
to the simulation of a small transverse area in
nucleus-nucleus collisions (the 
detailed study of nucleus-nucleus
collisions would require the introduction of 
a space-dependent $Q_s$ which we leave to a
future project).
These values of $\ell$ and $\ell_z$ correspond to
$\Delta_2^\mathrm{geom}=0$ ($\ell=\ell_z=0.5$ fm),
$\Delta_2^\mathrm{geom}=0.231$ ($\ell=1$ fm, $\ell_z=0.5$ fm)
and
$\Delta_2^\mathrm{geom}=0.306$ ($\ell=2$ fm, $\ell_z=0.5$ fm).

According to Eq.~\eqref{eq:spin_evol,_itisthereCL_ar}, 
spin also evolves thanks to the 
paramagnetic interaction with the
background gluon fields. 
We checked however that spin fluctuations
are not anisotropic, and in magnitude
they are always negligible in comparison
with the fluctuations of $\bm L$. 
We understand this 
in a simple way: if the gluon fields were time-independent and oriented
along the $z-$direction then the spin vector, $\bm S$,
would perform a precession motion, with
$S_x$ and $S_y$ rotating while
the $S_z$ component remaining constant: this could be 
a rough representation of the
motion of $\bm S$ in the very early stage
when the transverse fields are not 
large~\cite{Ruggieri:2017ioa}.
In this stage we would have 
$\langle S_x^2\rangle=\langle S_z^2\rangle=
\langle S_z^2\rangle=1/2$ (in natural units).
For larger times the transverse components
of the color-magnetic fields appear and the motion
of $\bm S$ becomes more complicated: nevertheless,
the condition $S_x^2 + S_y^2 + S_z^2=3/4$ is
satisfied at any time. Consequently, the components
of $\bm S$ cannot fluctuate as much as those of $\bm L$,
and the random variations of the
direction of the color-magnetic fields eventually
wash out the  anisotropy of the fluctuations
of $\bm S$. We confirmed this by a direct calculation.
Because of the isotropy and small size of the
spin fluctuations, we will not consider them in this
study and focus on the fluctuations of $\bm L$.

\section{Results\label{sec:results}}
\begin{figure}[t!]
	\centering
	\includegraphics[scale=.28]{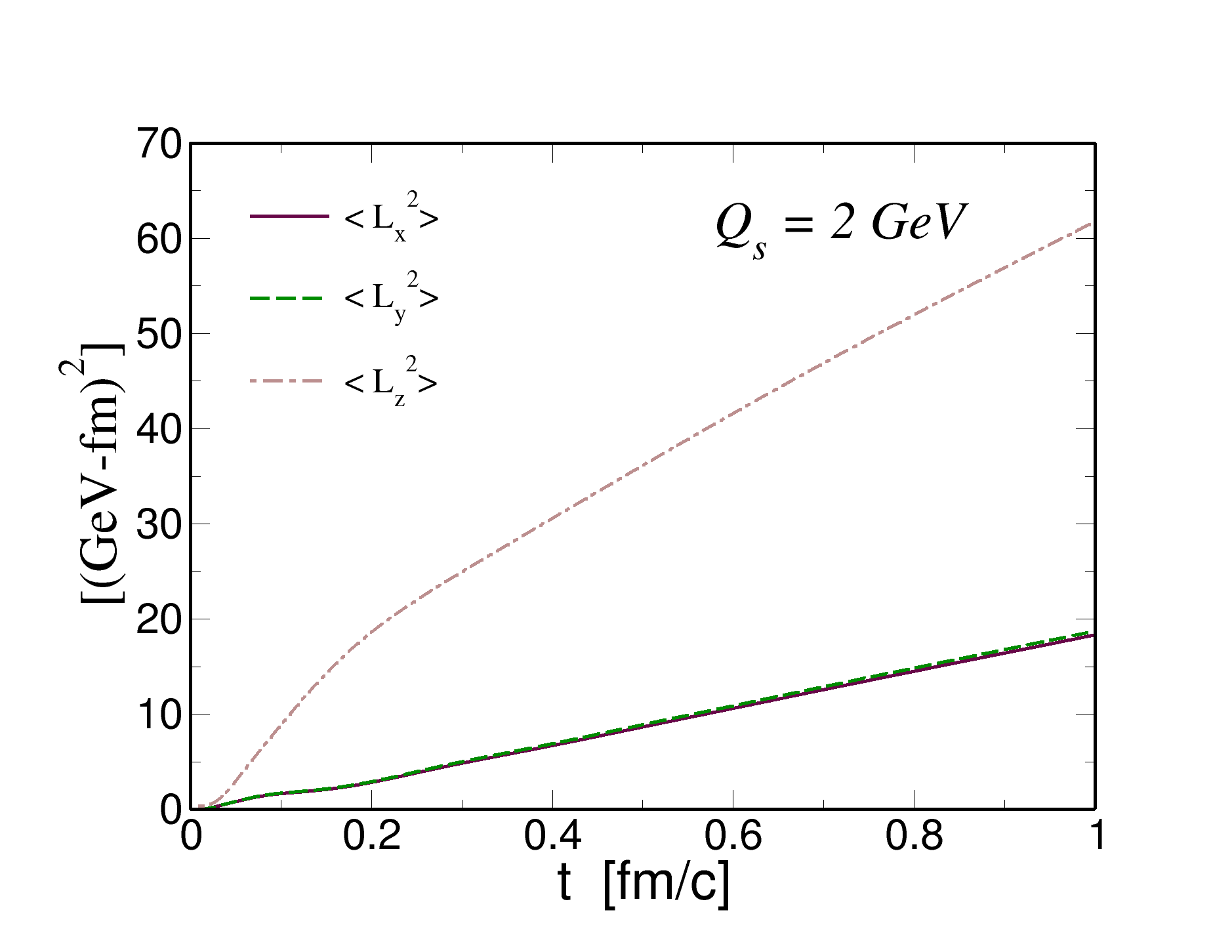}
	\caption{Orbital angular momentum components squared versus proper time for $m=10$ GeV,	for the	initial $p_T=0.5$ GeV.
	}
	\label{Fig:fig1_Lx2_Ly2_Lz2}
\end{figure}

In this section we summarize our results.
Numerical simulations have been performed
by fixing the value of $g\mu_A$ in Eq.~\eqref{eq:colorchargedensity_2} as
follows: we firstly
fix the saturation scale, $Q_s$,	
then the QCD coupling, $g$, by virtue of
the one-loop QCD $\alpha_s$ at the scale $Q_s$ so that	$g=\sqrt{4\pi\alpha_s}$. Hence, we get $g=g(Q_s)$. With the help of $g\mu_A \approx Q_s/0.6 g$ \cite{Lappi:2007ku}, fixing $g\mu_A$ is a straightforward task. 

HQs in our calculations are initialized as follows.
For each heavy quark, we create its companion anti-quark as well. They are initialized at time $\tau_{form} = 1/(2m)$ and immediately after their formation, they start interacting with the gluonic medium.
HQs are uniformly and randomly distributed in the
coordinate space to fill the entire box: for each
quark, its antiquark companion is initialized at the
same position. 
Then, 
the initial 
momentum distribution of heavy quarks is such that the transverse momentum $p_T$ is initialized like a $\delta$-distribution so that transverse components are distributed uniformly and longitudinal component vanishes. Along with this, we assume that each anti-quark carries the opposite momentum as that of its corresponding quark.
For both heavy quark and anti-quark, the set of $Q_a$ is initialized with uniform
probability on the sphere $Q^2 =Q_1^2 + Q_2^2 + Q_3^2 =1$, see  \cite{Liu:2020cpj} for more details.
The initial spins of quarks and anti-quarks are distributed uniformly on a sphere such that $S^2 =S_x^2 + S_y^2 + S_z^2$ with $S = \sqrt{3}/2$.
The initialisation of position and momentum of the colored probes help us to set up their initial orbital angular momentum components.

We perform numerical calculations implementing
a static box geometry for the evolving Glasma fields, leaving the more realistic case of the 
longitudinal expansion to a near future study. 
For each set of parameters, we perform $N_\mathrm{events}$	initializations and evolutions up to $t=1$ fm/c, then average the physical quantities over all the events. All the results correspond to $N_\mathrm{events}=100$, having verified that this is enough to achieve convergence.	

In this study we limit ourselves to consider
a very large, unphysical HQ mass, namely $m=10$ GeV,
in order to remain consistent with the non-relativistic
limit during the entire evolution: despite the 
academic flavor of this assumption, the evolution
of the angular momentum should be qualitatively 
comparable to that of HQs with physical masses,
since the anisotropic angular momentum 
fluctuations are related to the structure of the gluon
fields and not directly to the value of the HQ mass.
We also perform a set of calculations
with
$m=5$ GeV that is closer to the mass of the
beauty quarks, see Fig.~\ref{Fig:m5}.

In Fig.~\ref{Fig:moman} we plot the momentum
broadening of HQs in the evolving Glasma fields:
solid line denotes $\langle p_T^2\rangle$ while
dashed line stands for $\langle p_z^2\rangle$;
we remind that the $z-$direction coincides with the
initial direction of the gluon fields in the Glasma
(in realistic collisions, it is the flight direction
of the two colliding objects). 
The initial condition corresponds to 
$p_T=0.5$ GeV, $p_z=0.5$ GeV and we checked
that results are qualitatively unchanged
for different initial momenta.
We note that
the momentum broadening is anisotropic:
this is due to the anisotropy of the gluon 
fields \cite{Ruggieri:2018rzi,Ipp:2020nfu}.
As we mentioned in the previous section,
the anisotropic momentum broadening is 
vey important to generate anisotropic
angular momentum fluctuations.
 
In Fig.~\ref{Fig:fig1_Lx2_Ly2_Lz2}, we plot the square of the components of the angular momentum, namely $\langle L_x^2\rangle$, 
$\langle L_y^2\rangle$ and $\langle L_z^2\rangle$ 
for $Q_s=2$ GeV. 
For the system at hand 
$\langle L_x\rangle = \langle L_y\rangle =
\langle L_z\rangle=0$ during the entire evolution,
therefore the quantities shown in Fig.~\ref{Fig:fig1_Lx2_Ly2_Lz2}
correspond to the spreading of the components of $\bm L$,
similarly to the spreading of ordinary
momentum studied in the theory of
the Brownian motion, $\sigma_p=\langle
p^2 - \langle p\rangle^2\rangle$. 
We notice that $\langle L_z^2\rangle > 
\langle L_x^2\rangle,\langle L_y^2\rangle$.
Hence, the interaction of HQs with the gluon fields
produces anisotropic fluctuations of the
angular momentum.

\begin{figure}[t!]
	\begin{center}
	\includegraphics[scale=0.28]{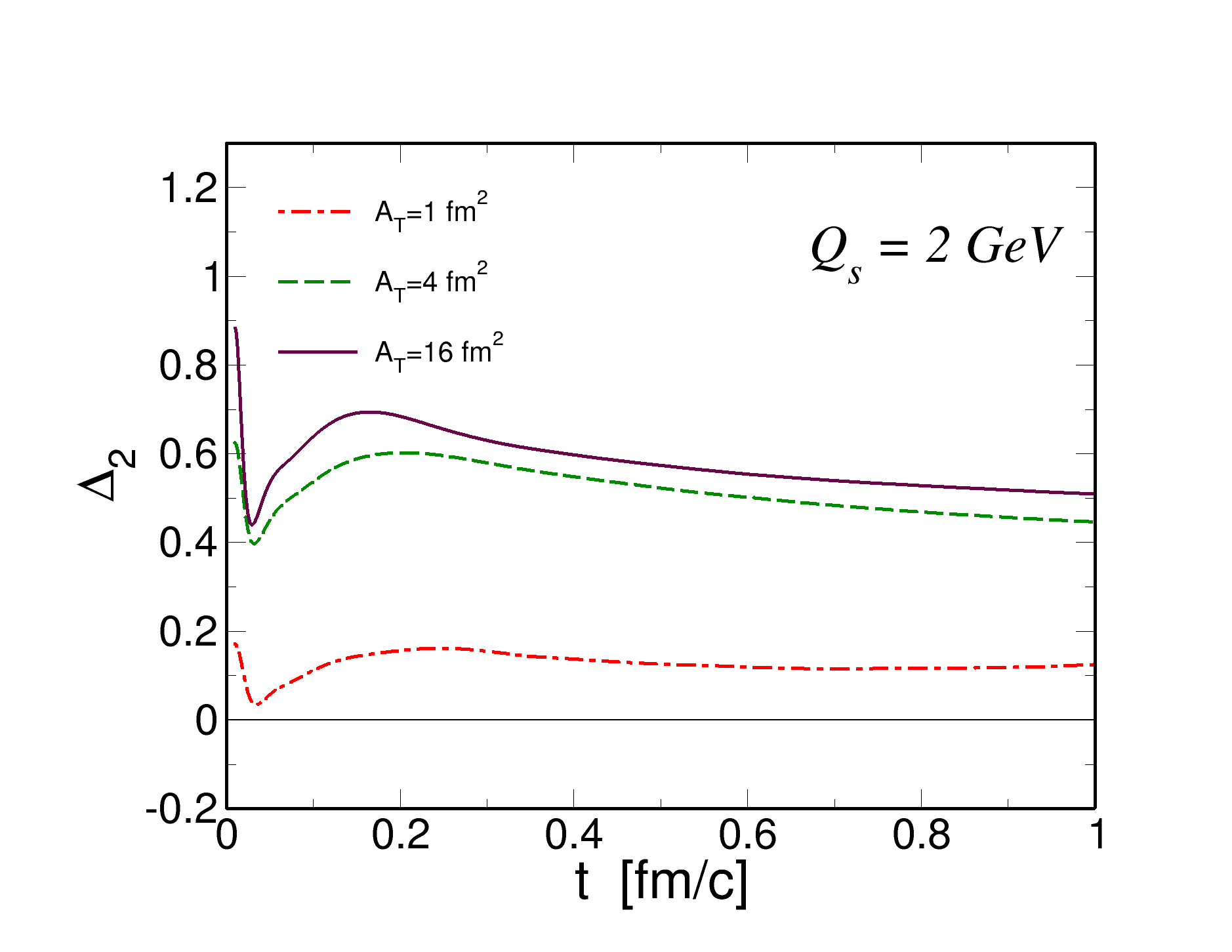}\\
		\includegraphics[scale=0.28]{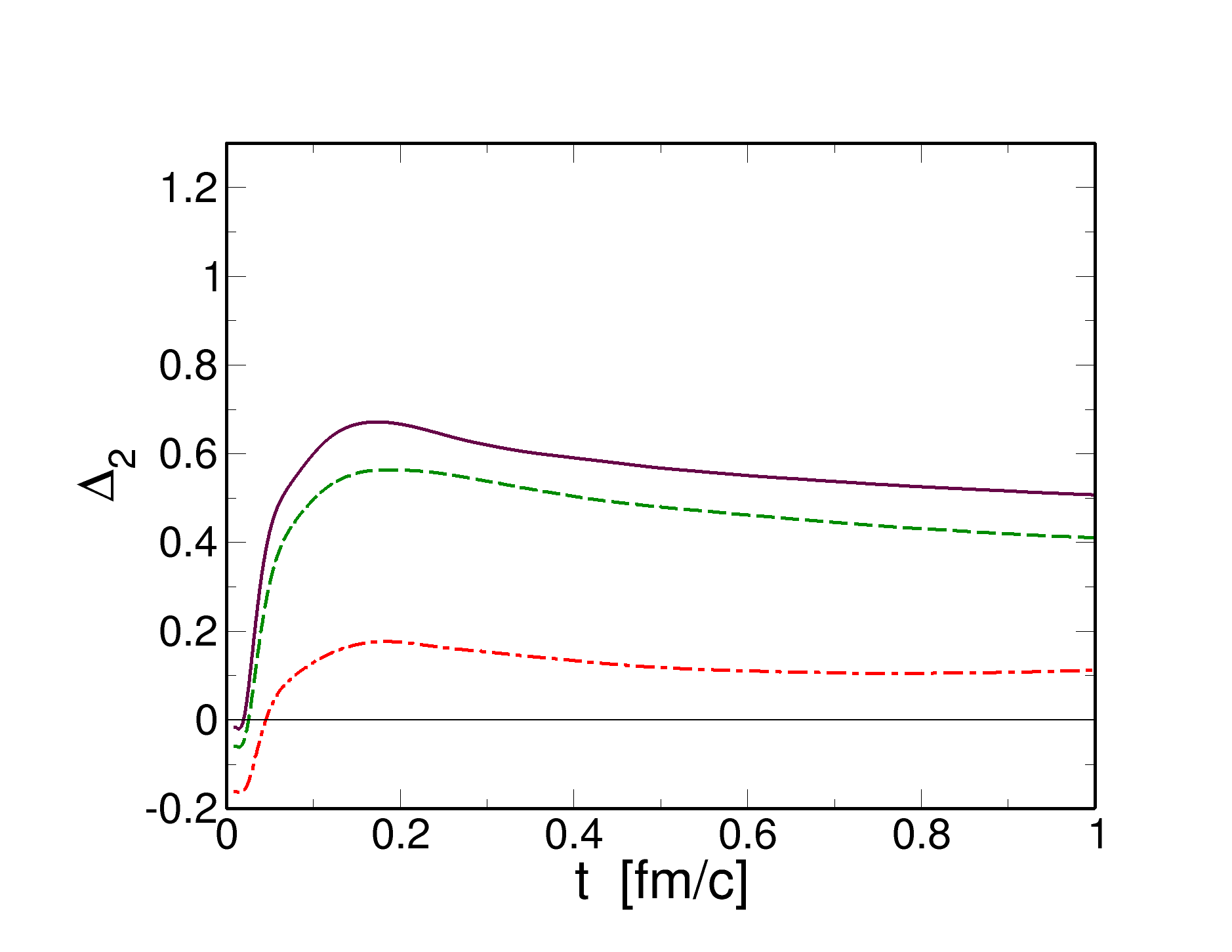}
	\end{center}
	\caption{Anisotropy parameter $\Delta_2$
	versus time, for $\ell=0.5$ fm ($A_T=1$ fm$^2$),
$\ell=1$ fm ($A_T=4$ fm$^2$) and 
$\ell=2$ fm ($A_T=16$ fm$^2$). Calculations 
correspond to $\ell_z=0.5$ fm. Upper and lower
panels
correspond to the initializations $p_T=0.5$ GeV, $p_z=0$ ($\eta=0$) 
and $p_T=0.5$ GeV, $|p_z|=0.5$ GeV 
respectively. Colors and line styles are the
same in the two panels.\label{Fig:d2}}
\end{figure} 

In Fig.~\ref{Fig:d2} we plot $\Delta_2$ versus time
for several values of $\ell$;
in the figure we defined the transverse area
of the box, 
$A_T=(2\ell)^2$.
We show results for $\ell=0.5$ fm ($A_T=1$ fm$^2$),
$\ell=1$ fm ($A_T=4$ fm$^2$) and 
$\ell=2$ fm ($A_T=16$ fm$^2$). Calculations 
correspond to $\ell_z=0.5$ fm. Upper and lower
panels
correspond to the initializations $p_T=0.5$ GeV, $p_z=0$ 
and $p_T=0.5$ GeV, $|p_z|=0.5$ GeV 
respectively. The saturation scale is
$Q_s=2$ GeV. 
With the initialization for $|p_z|=0.5$ GeV
we mean that $p_z=0.5$ GeV for $z>0$ and $p_z=-0.5$ GeV
for $z<0$: 
it is done to mimic 
the central rapidity region of a collision.

\begin{figure}[t!]
	\begin{center}
		\includegraphics[scale=0.28]{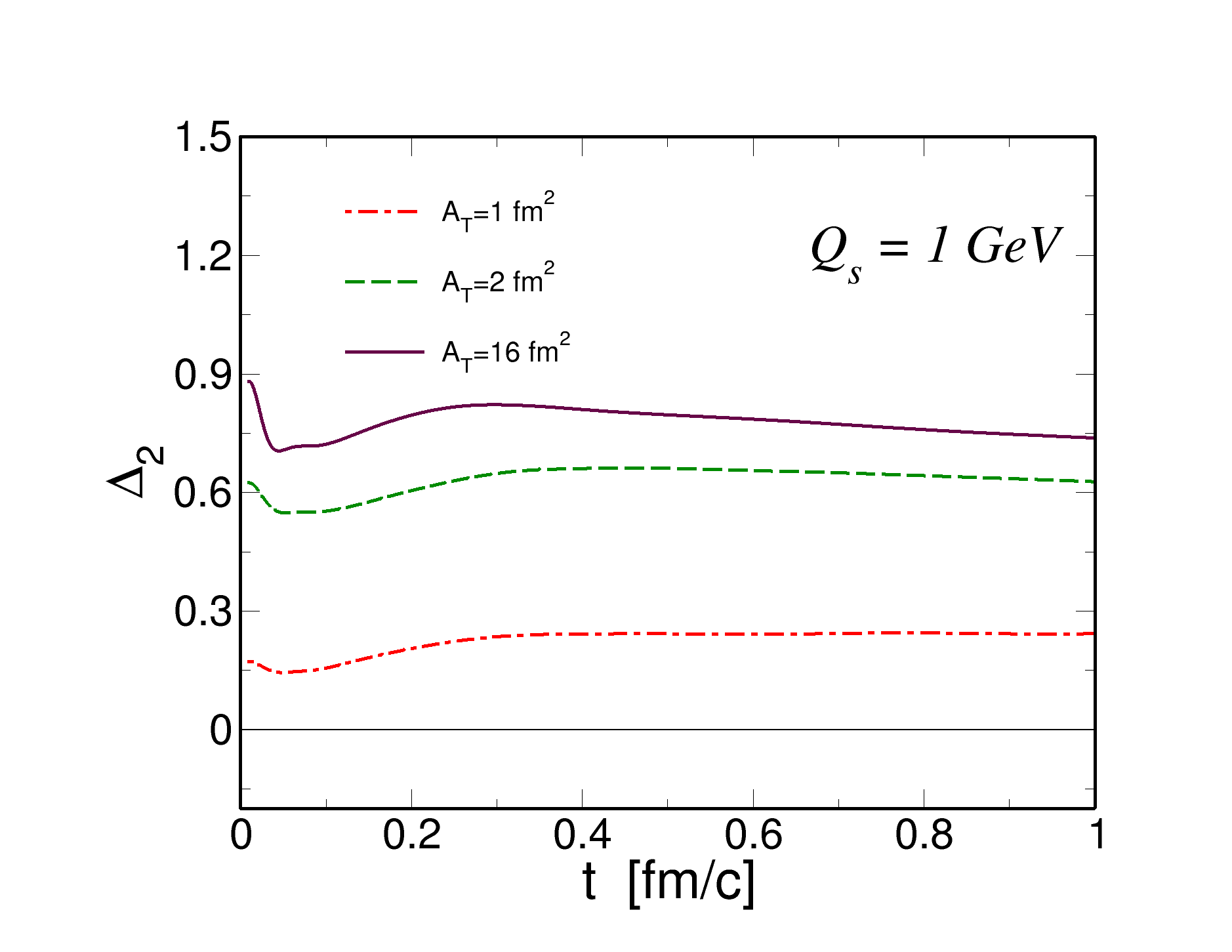}
	\end{center}
	\caption{$\Delta_2$ for $Q_s=1$ GeV. 
	Colors and line styles are the
	same of Fig.\ref{Fig:d2}.}
	\label{Fig:ap1}
\end{figure}

We note in Fig.~\ref{Fig:d2}
that in the case $p_z=0$,
besides a very quick transient, $\Delta_2$ increases
with time; within a time range that is
relevant for realistic collisions, that is up to
$t\approx 0.6$ fm/c, the anisotropy of the 
fluctuations of $\bm L$ lies within the $40\%$ and 
the $60\%$ in the cases of $\ell=1$ fm and $2$ fm
respectively. Changing the initialization
in $p_z$ simply changes a bit the short initial
transient, but at regime the results of this case
do not differ much with those obtained
for $p_z=0$. For $\ell=0.5$ fm $\Delta_2$ remains
smaller than the other two cases but it is still
sizeable. We repeat this analysis (limited to the
case $p_z=0$) for $Q_s= 1$ GeV and the result is
shown in Fig.~\ref{Fig:ap1}: we note that
changing the $Q_s$ does not lead to 
substantial changes
of $\Delta_2$. We thus expect that even using 
a distribution in $p_z$ should not change too much
the final value of $\Delta_2$.

We also note that the results shown
in Fig.~\ref{Fig:d2} are consistent with 
the discussion that lead to Eq.~\eqref{eq:an_78LN},
namely that increasing the transverse size
results in a $\Delta_2$ which is almost independent
on the size of the system. In fact, we note that 
while changing $\ell$ from $0.5$ fm to $1$ fm has a
substantial effect on $\Delta_2$, doubling again 
$\ell$ to $2$ fm does not change
$\Delta_2$ significantly. We finally checked that changing the
initial $p_T$ does not change qualitatively
the results.

\begin{figure}[t!]
	\begin{center}
		\includegraphics[scale=0.28]{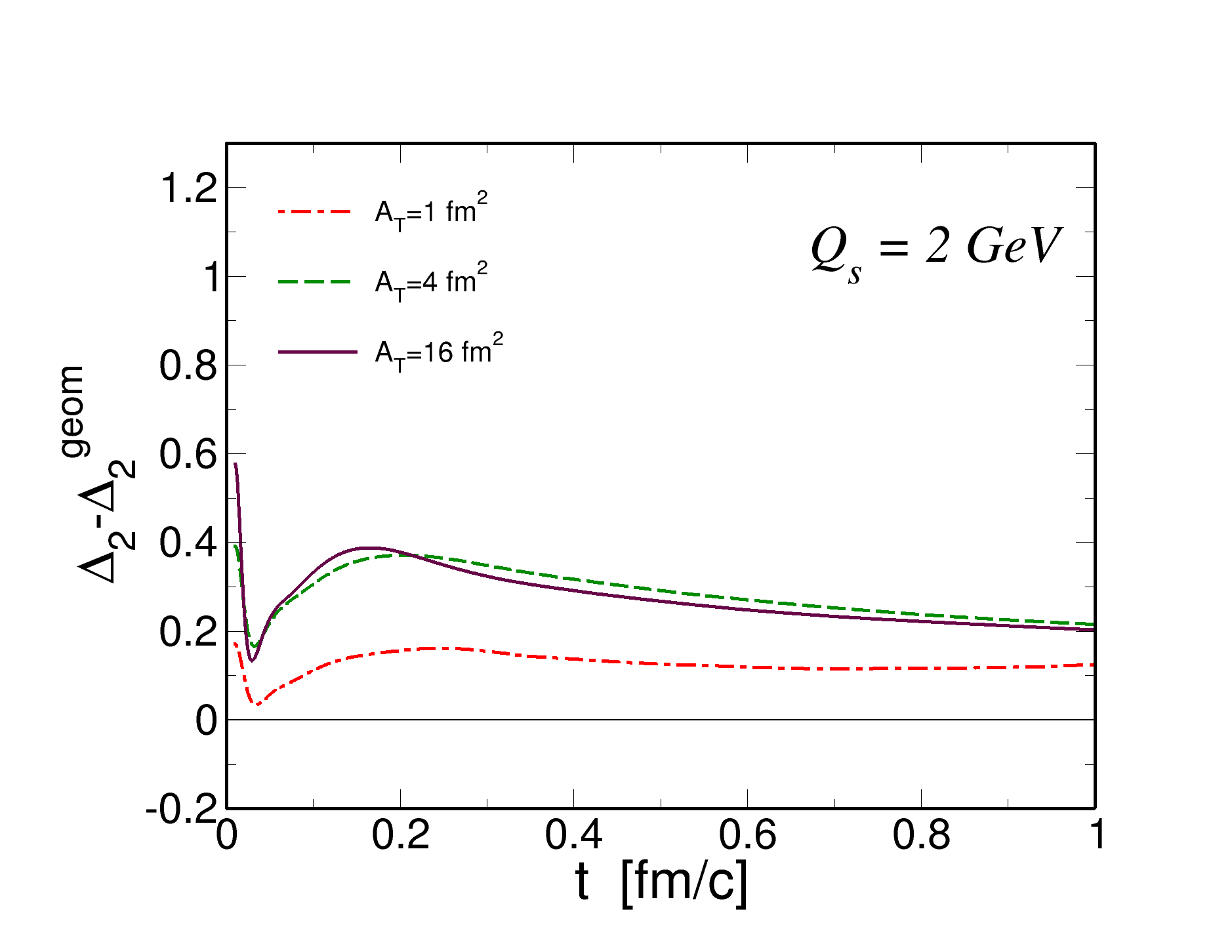}
	\end{center}
	\caption{$\Delta_2-\Delta_2^\mathrm{geom}$
		versus time.
		Colors and line styles are the
		same of Fig.\ref{Fig:d2}.\label{Fig:d2cl}}
\end{figure} 

As discussed in the previous section,
$\Delta_2$ potentially takes contribution
also from the anisotropic geometry of the fireball
and would be nonzero even if the momentum
broadening was isotropic. 
In order to prove 
that the results shown in Fig.~\ref{Fig:d2}
take a substantial contribution 
the anisotropic momentum broadening, we 
show 
$\Delta_2-\Delta_2^\mathrm{geom}$ 
in Fig.~\ref{Fig:d2cl} 
for the few values of the transverse area
already shown in Fig.~\ref{Fig:d2},  
and $\Delta_2^\mathrm{geom}$ is defined
in Eq.~\eqref{eq:lalonga}. We note that
$\Delta_2-\Delta_2^\mathrm{geom}$ 
is substantially different 
from $0$, leading us to 
conclude that $\Delta_2$
we show in Fig.~\ref{Fig:d2} takes a concrete
contribution from the anisotropy of the
momentum broadening. 

\begin{figure}[t!]
	\begin{center}
		\includegraphics[scale=0.28]{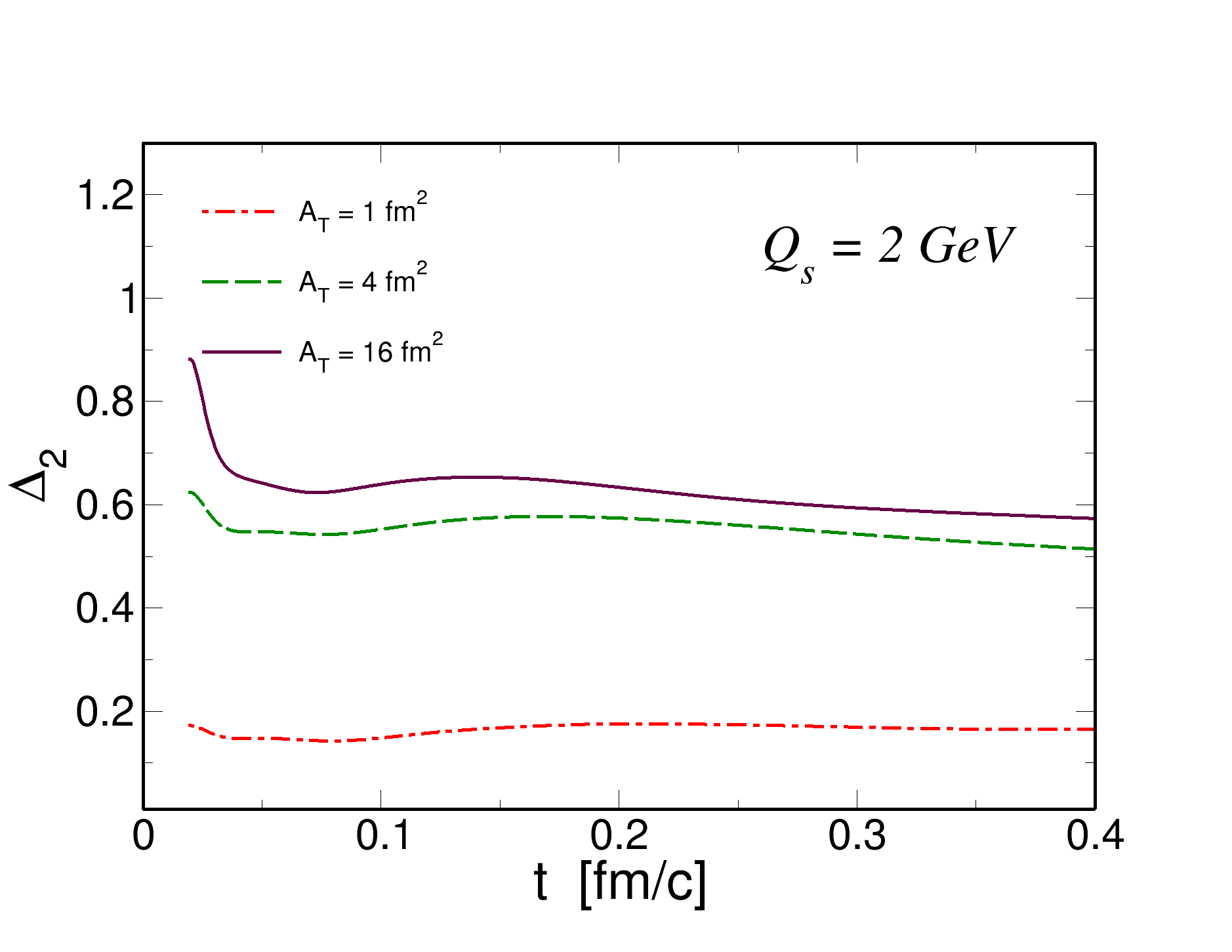}
	\end{center}
	\caption{Anisotropy parameter $\Delta_2$
		versus time, for $\ell=0.5$ fm ($A_T=1$ fm$^2$),
		$\ell=1$ fm ($A_T=4$ fm$^2$) and 
		$\ell=2$ fm ($A_T=16$ fm$^2$). Calculations 
		correspond to $\ell_z=0.5$ fm,
		$Q_s=2$ GeV and $m=5$ GeV. 
\label{Fig:m5}}
\end{figure} 

Although our calculations correspond to a large
$m$ limit, we can check whether our results 
qualitatively (and to some extent,
also quantitatively) stand for lower masses. 
To this end,
in Fig.~\ref{Fig:m5} we plot $\Delta_2$
versus time for $m=5$ GeV that roughly corresponds
to the mass of the beauty quark. We note that
both qualitatively and quantitatively the 
$\Delta_2$ agrees with that we found for 
a larger mass, suggesting that the anisotropy of
the fluctuations of $\bm L$ are quite robust.

As we discussed in the previous section,
we do not consider spin fluctuations here
because they are too small in magnitude compared
to the $\bm L-$fluctuations, and they remain
isotropic.
Moreover, fluctuations of $\bm S$ and
$\bm L$ are uncorrelated
(we confirmed this by direct calculations), 
therefore we can safely 
assume that the fluctuations of the total
angular momentum,
$\bm J=\bm L + \bm S$,
are
\begin{equation}
	\langle J^2_i\rangle =\langle
	( L_i + S_i)^2
	\rangle\approx \langle L_i^2 \rangle.
	\label{eq:totamg}
\end{equation}
As a consequence, $\Delta_2$
defined in Eq.~\eqref{eq:an_77} represents also the
anisotropy of the fluctuations of $J^2$,
hence our results predict anisotropic
fluctuations for the total angular momentum
of the HQs in the evolving Glasma fields.

\section{Conclusions and outlook\label{sec:conclusions}} 
We studied the fluctuations of angular momentum
of very massive quarks
in the pre-equilibrium stage
of relativistic nuclear collisions, in the background
of the evolving Glasma fields.
The heavy quark mass is taken to be $m=10$ GeV,
hence our model serves as a toy model of the
heavy quarks (HQs), charm and beauty, produced
in the collisions.
We solved
the non-relativistic kinetic equations in which the HQs
diffuse in the background gluon fields; the evolution
of the latters is obtained by solving the Yang-Mills
equations with the Glasma initial condition.
The implementation of the realistic HQs masses,
namely charm and beauty, require the use of
relativistic equations and these will be the subject
of a future study.

We found that the fluctuations of the angular momentum,
$\bm L$, of HQs are anisotropic, and we explained this
in terms of the anisotropy of the gluon fields
which in turn results in the anisotropic fluctuations
of momentum. We introduced an anisotropy parameter,
$\Delta_2$ in Eq.~\eqref{eq:an_77}, that measures the
anisotropy of the fluctuations along the $z-$ and
$x-$directions and
that we showed
in Fig.~\ref{Fig:d2}. We found that 
the anisotropy develops quite
early and can be as large as the $60\%$ for 
larger systems, remaining of about the $40\%$ for
transverse areas of the order of the proton size.
We checked that the qualitative picture is unchanged
by changing the initialization in momentum space
as well the value of the saturation scale, $Q_s$.
Moreover, we computed the evolution of spin
fluctuations but we found no sign of 
anisotropy in these fluctuations: it is likely
that this happens as a combination of the random
fluctuations of the color-magnetic fields as well
as of the large value of $m$ (paramagnetic coupling
is suppressed by a power of $m$).

Considering that the fluctuations of $L^2$ are way
larger than those of $S^2$, and that $\bm L$ and
$\bm S$ are uncorrelated, we can safely assume that
the fluctuations of the total angular momentum 
coincides with those of the components of $\bm L$, that 
is $\langle J_i^2\rangle = \langle L_i^2\rangle$
for $i=x,y,z$. Therefore, the results of our study
suggest that the fluctuations of the 
total angular momentum of HQs in the evolving
Glasma fields are anisotropic.

We would like to stress that even though
our results were obtained by using toy heavy quarks,
we do not expect that
the qualitative behavior of the quantities we 
computed will change when realistic quark masses
will be implemented: as a matter of fact,
the anisotropy of the fluctuations of the
angular momentum
 depends on the structure of the
gluon fields in the evolving Glasma and not on the
value of the quark mass, therefore even lowering the
value of the latter should not change drastically,
or at least qualitatively,
our results.

There are several ways in which this study can be
improved. First of all, we would like to 
formulate the problem relativistically, 
in order to make calculations with the
realistic values of the 
masses of charm and beauty quarks.
The equations for this case
already exist in the literature~\cite{Heinz:1984yq}
and can be
implemented within our scheme. 
Turning to the relativistic equations
will also trigger interesting,
purely relativistic effects~\cite{Becattini:2022zvf}
that might give correlations
between $\bm L$ and $\bm S$.
In fact, within our scheme we already have
a direct coupling of $\bm S$ to $\bm P$,
see Eq.~\eqref{eq:acca_91_3naMF_2_ar},
and hence to $\bm L$,
but this coupling is suppressed by
a power of the HQ mass and gives 
a negligible contribution to the evolution
of the physical quantities:
it is likely that in the relativistic
calculation that term gives a sizeable effect.
Moreover,
in the relativistic case our Eq.~\eqref{eq:spin_evol,_itisthereCL_ar} 
has to be replaced by a BMT-like equation~\cite{Bargmann:1959gz,Heinz:1984yq},
in which the spin couples to the
$4-$momentum as well as
to the inhomogeneities of the gauge fields;
in the full relativistic calculation
this coupling, which is of the order
of $1/m^2$, would join the aforementioned couplings
already present in the non-relativistic limit,
and could 
also strengthen the
correlation between
 $\bm L$ and $\bm S$:
as a result the anisotropy
of the fluctuations of $\bm L$ that we discussed
could be transferred to the fluctuations of $\bm S$. 
In addition to this, 
it would be interesting to prepare
a rapidity-dependent initialization
of the gluon fields,
similarly to what
already done in relativistic hydrodynamics
and kinetic theory~\cite{Becattini:2015ska,Oliva:2020doe}
that would allow us to have a nonzero average 
angular momentum of the bulk and of the
HQs.
Furthermore, in order
to make more concrete phenomenological predictions,
one should prepare a more realistic initial condition
in which the $Q_s$ fluctuates on the transverse plane,
then attach the evolution 
of the HQs in the Glasma fields
to that in the QGP droplets and eventually
to hadronization: this is is quite an 
ambitious project and will be the subject of future studies.


\begin{acknowledgements}
M.R. acknowledges John Petrucci for inspiration. 
Lucia Oliva 
is acknowledged for the numerous discussions
on the topics of the research presented here.
S.K.D. and M.R. acknowledge the support by the National Science Foundation of China (Grant Nos. 11805087 and 11875153). S.K.D. acknowledges the support from DAE-BRNS, India, Project No. 57/14/02/2021-BRNS.
\end{acknowledgements}



\begin{thebibliography}{50}

\bibitem{Shuryak:2004cy}
E.~V.~Shuryak,
Nucl. Phys. A \textbf{750}, 64-83 (2005)

\bibitem{Jacak:2012dx}
B.~V.~Jacak and B.~Muller,
Science \textbf{337}, 310-314 (2012)



\bibitem{McLerran:1993ni} 
L.~D.~McLerran and R.~Venugopalan,
Phys.\ Rev.\ D {\bf 49}, 2233 (1994)

\bibitem{McLerran:1993ka} 
L.~D.~McLerran and R.~Venugopalan,
Phys.\ Rev.\ D {\bf 49}, 3352 (1994)

\bibitem{McLerran:1994vd} 
L.~D.~McLerran and R.~Venugopalan,
Phys.\ Rev.\ D {\bf 50}, 2225 (1994)

\bibitem{Gelis:2010nm} 
F.~Gelis, E.~Iancu, J.~Jalilian-Marian and R.~Venugopalan,
Ann.\ Rev.\ Nucl.\ Part.\ Sci.\  {\bf 60}, 463 (2010).


\bibitem{Iancu:2003xm} 
E.~Iancu and R.~Venugopalan,
In *Hwa, R.C. (ed.) et al.: Quark gluon plasma* 249-3363.



\bibitem{McLerran:2008es} 
L.~McLerran,
arXiv:0812.4989 [hep-ph];
hep-ph/0402137.

\bibitem{Gelis:2012ri} 
F.~Gelis,
Int.\ J.\ Mod.\ Phys.\ A {\bf 28}, 1330001 (2013)




\bibitem{Kovner:1995ja} 
A.~Kovner, L.~D.~McLerran and H.~Weigert,
Phys.\ Rev.\ D {\bf 52}, 6231 (1995)

\bibitem{Kovner:1995ts} 
A.~Kovner, L.~D.~McLerran and H.~Weigert, Phys.\ Rev.\ D {\bf 52}, 3809 (1995)

\bibitem{Gyulassy:1997vt} 
M.~Gyulassy and L.~D.~McLerran,
Phys.\ Rev.\ C {\bf 56}, 2219 (1997)

\bibitem{Lappi:2006fp} 
T.~Lappi and L.~McLerran,
Nucl.\ Phys.\ A {\bf 772}, 200 (2006)



\bibitem{Krasnitz:2003jw} 
A.~Krasnitz, Y.~Nara and R.~Venugopalan,
Nucl.\ Phys.\ A {\bf 727}, 427 (2003)


\bibitem{Fukushima:2006ax} 
K.~Fukushima, F.~Gelis and L.~McLerran,
Nucl.\ Phys.\ A {\bf 786}, 107 (2007)


\bibitem{Fujii:2008km} 
H.~Fujii, K.~Fukushima and Y.~Hidaka,
Phys.\ Rev.\ C {\bf 79}, 024909 (2009)


\bibitem{Fukushima:2013dma} 
K.~Fukushima,
Phys.\ Rev.\ C {\bf 89}, no. 2, 024907 (2014)



\bibitem{Romatschke:2005pm} 
P.~Romatschke and R.~Venugopalan,
Phys.\ Rev.\ Lett.\  {\bf 96}, 062302 (2006)


\bibitem{Romatschke:2006nk} 
P.~Romatschke and R.~Venugopalan,
Phys.\ Rev.\ D {\bf 74}, 045011 (2006)


\bibitem{Fukushima:2011nq} 
K.~Fukushima and F.~Gelis,
Nucl.\ Phys.\ A {\bf 874}, 108 (2012).












\bibitem{Prino:2016cni} 
F.~Prino and R.~Rapp,
J.\ Phys.\ G {\bf 43}, no. 9, 093002 (2016)


\bibitem{Andronic:2015wma} 
A.~Andronic {\it et al.},
Eur.\ Phys.\ J.\ C {\bf 76}, no. 3, 107 (2016)



\bibitem{Rapp:2018qla}
R.~Rapp, P.~B.~Gossiaux, A.~Andronic, R.~Averbeck, S.~Masciocchi, A.~Beraudo, E.~Bratkovskaya, P.~Braun-Munzinger, S.~Cao and A.~Dainese, \textit{et al.}
Nucl. Phys. A \textbf{979} (2018), 21-86


\bibitem{Cao:2018ews}
S.~Cao, G.~Coci, S.~K.~Das, W.~Ke, S.~Y.~F.~Liu, S.~Plumari, T.~Song, Y.~Xu, J.~Aichelin and S.~Bass, \textit{et al.}
Phys. Rev. C \textbf{99}, no.5, 054907 (2019)

\bibitem{Aarts:2016hap} 
G.~Aarts {\it et al.},
Eur.\ Phys.\ J.\ A {\bf 53}, no. 5, 93 (2017)


\bibitem{Greco:2017rro} 
V.~Greco,
Nucl.\ Phys.\ A {\bf 967}, 200 (2017).


\bibitem{Dong:2019unq}
X.~Dong and V.~Greco,
Prog. Part. Nucl. Phys. \textbf{104} (2019), 97-141


\bibitem{Xu:2018gux}
Y.~Xu, S.~A.~Bass, P.~Moreau, T.~Song, M.~Nahrgang, E.~Bratkovskaya, P.~Gossiaux, J.~Aichelin, S.~Cao and V.~Greco, \textit{et al.}
Phys. Rev. C \textbf{99} (2019) no.1, 014902





\bibitem{Moore:2004tg}
G.~D.~Moore and D.~Teaney,
Phys.\ Rev.\ C {\bf 71}, 064904 (2005)


\bibitem{vanHees:2005wb}
H.~van Hees, V.~Greco and R.~Rapp,
Phys.\ Rev.\ C {\bf 73}, 034913 (2006)


\bibitem{vanHees:2007me}
H.~van Hees, M.~Mannarelli, V.~Greco and R.~Rapp,
Phys.\ Rev.\ Lett.\  {\bf 100}, 192301 (2008)

\bibitem{Gossiaux:2008jv}
P.~B.~Gossiaux and J.~Aichelin,
Phys.\ Rev.\ C {\bf 78}, 014904 (2008)

\bibitem{He:2011qa}
M.~He, R.~J.~Fries and R.~Rapp,
Phys.\ Rev.\ C {\bf 86}, 014903 (2012)

\bibitem{Prakash:2021lwt}
J.~Prakash, M.~Kurian, S.~K.~Das and V.~Chandra,
Phys. Rev. D \textbf{103} (2021) no.9, 094009



\bibitem{Song:2015sfa}
T.~Song, H.~Berrehrah, D.~Cabrera, J.~M.~Torres-Rincon, L.~Tolos, W.~Cassing and E.~Bratkovskaya,
Phys.\ Rev.\ C {\bf 92}, no. 1, 014910 (2015)




\bibitem{Alberico:2011zy}
W.~M.~Alberico, A.~Beraudo, A.~De Pace, A.~Molinari, M.~Monteno, M.~Nardi and F.~Prino,
Eur.\ Phys.\ J.\ C {\bf 71}, 1666 (2011)



\bibitem{Lang:2012cx}
T.~Lang, H.~van Hees, J.~Steinheimer, G.~Inghirami and M.~Bleicher,
Phys.\ Rev.\ C {\bf 93}, no. 1, 014901 (2016)



\bibitem{Das:2015ana} 
S.~K.~Das, F.~Scardina, S.~Plumari and V.~Greco,
Phys.\ Lett.\ B {\bf 747}, 260 (2015)


\bibitem{Xu:2017obm}
Y.~Xu, J.~E.~Bernhard, S.~A.~Bass, M.~Nahrgang and S.~Cao,
Phys.\ Rev.\ C {\bf 97}, no. 1, 014907 (2018)

\bibitem{Cao:2016gvr}
S.~Cao, T.~Luo, G.~Y.~Qin and X.~N.~Wang,
Phys.\ Rev.\ C {\bf 94}, no. 1, 014909 (2016)








\bibitem{Das:2016cwd} 
S.~K.~Das, S.~Plumari, S.~Chatterjee, J.~Alam, F.~Scardina and V.~Greco,
Phys.\ Lett.\ B {\bf 768}, 260 (2017)



\bibitem{Das:2017dsh} 
S.~K.~Das, M.~Ruggieri, F.~Scardina, S.~Plumari and V.~Greco,
J.\ Phys.\ G {\bf 44}, no. 9, 095102 (2017)



\bibitem{Das:2015aga} 
S.~K.~Das, M.~Ruggieri, S.~Mazumder, V.~Greco and J.~e.~Alam,
J.\ Phys.\ G {\bf 42}, no. 9, 095108 (2015)


\bibitem{Song:2019cqz}
T.~Song, P.~Moreau, J.~Aichelin and E.~Bratkovskaya,
Phys. Rev. C \textbf{101} (2020) no.4, 044901


\bibitem{Beraudo:2015wsd} 
A.~Beraudo, A.~De Pace, M.~Monteno, M.~Nardi and F.~Prino,
JHEP {\bf 1603}, 123 (2016)





\bibitem{Das:2013kea} 
S.~K.~Das, F.~Scardina, S.~Plumari and V.~Greco,
Phys.\ Rev.\ C {\bf 90}, 044901 (2014)

\bibitem{Berrehrah:2013mua}
H.~Berrehrah, E.~Bratkovskaya, W.~Cassing, P.~B.~Gossiaux, J.~Aichelin and M.~Bleicher,
Phys. Rev. C \textbf{89}, no.5, 054901 (2014)


\bibitem{Scardina:2017ipo}
F.~Scardina, S.~K.~Das, V.~Minissale, S.~Plumari and V.~Greco,
Phys. Rev. C \textbf{96}, no.4, 044905 (2017)


\bibitem{Mrowczynski:2017kso} 
S.~Mrowczynski,
Eur.\ Phys.\ J.\ A {\bf 54}, no. 3, 43 (2018)
 
 
\bibitem{Ruggieri:2018rzi}
M.~Ruggieri and S.~K.~Das,
Phys. Rev. D \textbf{98}, no.9, 094024 (2018)



\bibitem{Sun:2019fud}
Y.~Sun, G.~Coci, S.~K.~Das, S.~Plumari, M.~Ruggieri and V.~Greco,
Phys. Lett. B \textbf{798}, 134933 (2019)


\bibitem{Liu:2019lac}
J.~H.~Liu, S.~Plumari, S.~K.~Das, V.~Greco and M.~Ruggieri,
Phys. Rev. C \textbf{102}, no.4, 044902 (2020)


\bibitem{Boguslavski:2020tqz}
K.~Boguslavski, A.~Kurkela, T.~Lappi and J.~Peuron,

\bibitem{Liu:2020cpj}
J.~H.~Liu, S.~K.~Das, V.~Greco and M.~Ruggieri,
Phys. Rev. D \textbf{103} (2021) no.3, 034029


\bibitem{Khowal:2021zoo}
P.~Khowal, S.~K.~Das, L.~Oliva and M.~Ruggieri, Eur. Phys. J. Plus \textbf{137} (2022) no.3, 307


\bibitem{Ipp:2020nfu}
A.~Ipp, D.~I.~M\"uller and D.~Schuh,
Phys. Lett. B \textbf{810}, 135810 (2020)


\bibitem{Kunihiro:2010tg}
T.~Kunihiro, B.~Muller, A.~Ohnishi, A.~Schafer, T.~T.~Takahashi and A.~Yamamoto,
Phys. Rev. D \textbf{82}, 114015 (2010)


\bibitem{Heinz:1984yq}
U.~W.~Heinz,
Annals Phys. \textbf{161}, 48 (1985)


\bibitem{Bargmann:1959gz}
V.~Bargmann, L.~Michel and V.~L.~Telegdi,
Phys. Rev. Lett. \textbf{2}, 435-436 (1959)


\bibitem{Lappi:2007ku}
T.~Lappi,
Eur. Phys. J. C \textbf{55}, 285-292 (2008)

\bibitem{Ruggieri:2017ioa}
M.~Ruggieri, L.~Oliva, G.~X.~Peng and V.~Greco,
Phys. Rev. D \textbf{97}, no.7, 076004 (2018)

\bibitem{Becattini:2022zvf}
F.~Becattini,
[arXiv:2204.01144 [nucl-th]].


\cite{Becattini:2015ska}
\bibitem{Becattini:2015ska}
F.~Becattini, G.~Inghirami, V.~Rolando, A.~Beraudo, L.~Del Zanna, A.~De Pace, M.~Nardi, G.~Pagliara and V.~Chandra,
Eur. Phys. J. C \textbf{75}, no.9, 406 (2015)
[erratum: Eur. Phys. J. C \textbf{78}, no.5, 354 (2018)]

\bibitem{Oliva:2020doe}
L.~Oliva, S.~Plumari and V.~Greco,
JHEP \textbf{05}, 034 (2021)
































\end{thebibliography}
\end{document}